\documentclass[prb,twocolumn]{revtex4-1}
\usepackage{graphicx}
\usepackage{float}

\usepackage{amssymb}
\usepackage{amsmath}
\usepackage{dcolumn}
\usepackage{colortbl}

\definecolor{LightGray}{gray}{0.85}

\bibpunct{[}{]}{,}{n}{}{}

\begin{document}


\title{Finite-time scaling at the Anderson transition for vibrations in solids}


\author{Y.M. Beltukov}
\email{ybeltukov@gmail.com}
\affiliation{Ioffe Institute, 194021 St. Petersburg, Russia}
\author{S.E. Skipetrov}
\email{sergey.skipetrov@lpmmc.cnrs.fr}
\affiliation{Univ. Grenoble Alpes, CNRS, LPMMC, 38000 Grenoble, France}

\date{\today}

\begin{abstract}
A model in which a three-dimensional elastic medium is represented by a network of identical masses connected by springs of random strengths and allowed to vibrate only along a selected axis of the reference frame, exhibits an Anderson localization transition. To study this transition, we assume that the dynamical matrix of the network is given by a product of a sparse random matrix with real, independent, Gaussian-distributed non-zero entries and its transpose. A finite-time scaling analysis of system's response to an initial excitation allows us to estimate the critical parameters of the localization transition. The critical exponent is found to be $\nu = 1.57 \pm 0.02$ in agreement with previous studies of Anderson transition belonging to the three-dimensional orthogonal universality class.
\end{abstract}

\maketitle

\section{Introduction}
\label{sec_intro}

The phenomenon of Anderson localization---a halt of wave transport by disorder---has been intensively studied for almost 60 years since its discovery by P.W. Anderson in 1958 \cite{anderson58,lagendijk09,abrahams10}. First considered as a quantum effect specific for electrons in disordered solids at low temperatures \cite{anderson58}, it was later given a simpler and more appropriate interpretation as an interference effect that can take place for any wave in the presence of a sufficiently strong disorder \cite{john83,john84,anderson85}. Even though quantum systems continue to serve as a playground to study Anderson localization, with, in particular, impressive successes achieved using cold atoms \cite{billy08,chabe08,jendr12}, an increasing amount of new knowledge in this field is now acquired from experiments with ``classical'' waves, such as microwaves \cite{chabanov00,chabanov01,shi14}, light \cite{wiersma97,storzer06,vanderbeek12,sperling13,sperling16}, or sound \cite{hu08,hilde14,aubry14,cobus16}. The latter allow for easier and less expensive experiments requiring neither low temperatures (because classical waves can have long coherence lengths at room temperatures) nor small sample sizes (because coherence lengths of several meters or even longer can be readily achieved). Even if these experiments come, of course, with their own difficulties, such as, e.g., the absorption of wave energy by the disordered sample \cite{scheffold99} or inelastic scattering effects \cite{scheffold13}, they have an undeniable advantage of allowing for more versatile studies of Anderson localization because measurements of scattered wave fields can be readily performed with spatial, angular, temporal, and frequency resolution.

Mechanical vibrations in elastic solids constitute an example of wave excitations that unequivocally demonstrate the classical nature of the Anderson localization phenomenon. Indeed, the localization of vibrational modes can be fully understood based on a model derived from the Newton equation, which is the ``most classical'' equation in physics. Numerous experimental studies of localization of vibrational modes exist in one \cite{he86,flores13} and two \cite{weaver90,lobkis08} dimensions where theoretical analysis, both analytic and numerical, is also easier to carry out \cite{sheng86,weaver94,sigalas05}. In contrast, experiments clearly demonstrating Anderson localization of vibrations in three-dimensional (3D) disordered solids are scarce and relatively recent \cite{hu08,hilde14,aubry14,cobus16}; the analytic theory is hard to develop and often relies on crude approximations \cite{john83,kirk85,condat87,photiadis17}, and numerics is time- and computer-resource consuming \cite{sheng91,sheng94,sepehrinia08,pinski12,pinski12a,amir13}. Nevertheless, the 3D case is the most interesting one because, in contrast to lower dimensions where all modes are localized whatever the disorder strength \cite{loc2d}, a localization transition is expected in 3D \cite{abrahams79,evers08}. Not only this transition has a number of interesting critical properties (critical exponents, multifractality of critical modes, etc.) but it also constitutes a link between different disordered physical systems (electrons in disordered solids, cold atoms in random potentials, light in disordered dielectrics, vibrations in solids, etc.), which are all expected to exhibit the same, \textit{universal} critical behavior at the Anderson transition despite their differences \cite{evers08}.

In the present paper we propose a numerical approach to the study of Anderson localization of vibrations in disordered solids. It differs from the previous ones in two important aspects. First, it is based on the use of sparse random matrices with independent elements obeying Gaussian statistics. This makes the corresponding theoretical model potentially suitable for analytic treatments. Second, the critical properties of the localization transition are determined using a procedure of finite-\textit{time} scaling, to be contrasted with the more common procedure of finite-size scaling \cite{mackinnon81,pichard81,shklovskii93} (see Ref.\ \onlinecite{slevin14} for a review). The latter has gained of lot of popularity in the studies of Anderson localization via numerical simulations but its basic principles are difficult or even impossible to put in practice in a realistic experiment, where preparing a set of disordered samples of different sizes but with identical disorder statistics and strengths may represent an insurmountable difficulty. In contrast, the finite-time scaling approach can be applied to experimental data obtained on a single disordered sample provided that time-dependent measurements are performed in a sufficiently wide temporal range. It has been recently used to study Anderson transition in a cold-atom system \cite{chabe08,lemarie09} and has a potential of being applied to the time-dependent experimental data of Refs.\ \cite{hu08,cobus16}. Thus, our work represents a useful step towards an experimental realization of finite-time scaling for mechanical vibrations in disordered solids by demonstrating the single-parameter scaling in a model system describing mechanical vibrations and yielding the value of the critical exponent of the localization transition which is compatible with the value obtained previously by the finite-size scaling approach. In addition, our computational method is quite efficient and may compete, in terms of calculational efficiency, with the now standard transfer-matrix techniques used in finite-size scaling calculations. Practical applications of our results may lie in the field of thermal transport in amorphous or other disordered materials, to which vibrational modes (phonons) contribute \cite{sheng91,chaud10}.

\section{The model}
\label{sec_model}

Consider a 3D cubic lattice of $N$ identical point masses $m=1$ (atoms) connected by harmonic springs. The equations of motion for displacements ${\bf u}_i(t) = {\bf r}_i(t) - {\bf R}_i$ of atoms from equilibrium positions ${\bf R}_i$ can be written as~\cite{Beltukov-2013}
\begin{equation}
    \ddot{u}_i^\alpha(t) = -\sum_{j\beta} M_{ij}^{\alpha\beta}u_j^\beta(t),  \label{eq:newton}
\end{equation}
where $M_{ij}^{\alpha\beta}$ are elements of the dynamical matrix ${\hat M}$. In the following, we apply the scalar approximation, in which indices $\alpha$ and $\beta$ of Cartesian coordinates are omitted. Simply speaking, in the scalar approximation atoms are allowed to move only along a selected axis of the reference frame. The displacements ${\bf u}_i$ then become scalar $u_i$.

Eigenvalues of the dynamical matrix ${\hat M} = \{ M_{ij} \}_{N \times N}$ are squared eigenfrequencies $\omega_n^2$. Thus, the dynamical matrix of a mechanically stable system is positive semidefinite, which is possible if and only if the dynamical matrix can be represented as ${\hat M} = {\hat A} {\hat A}^T$~\cite{Beltukov-2013}. Off-diagonal matrix elements $A_{ij}$ ($i \ne j$), describing interactions between nearest-neighbor atoms, are assumed to be independent, zero-mean Gaussian variables with identical variances $\Omega^2$. $A_{ij} = 0$ for atoms which are not nearest neighbors. The diagonal elements $A_{ii}$ are obtained by a sum rule $A_{ii} = -\sum_{j\ne i} A_{ji}$, which ensures that the total potential energy is invariant upon translation of the system as a whole. Hence, the random matrix ${\hat A}$ has $7N$ non-zero elements (coupling with 6 nearest neighbors for each of $N$ atoms, plus $N$ diagonal elements), of which only $6N$ are statistically independent. We put $\Omega=1$, which fixes the frequency and time units.

In the framework of the model described above, we run a series of calculations for a large system ($N = 200^3$) starting with a homogeneously excited left half of the sample $x < 0$. Initial velocities $\dot{u}_i(0)$ are taken to be Gaussian random numbers if $x_i<0$ and zeros otherwise. Initial displacements $u_i(0)$ are set to zero. We calculate atomic displacements $u_i(t)$ and atomic velocities $\dot{u}_i(t)$ using Verlet integration and observe spreading of the vibrational energy from the left ($x<0$) to the right ($x>0$) half of the system as time increases. This process is a superposition of many individual harmonic processes, corresponding to different frequencies $\omega$. Previous calculations have shown that spreading of vibrational energy with time slows down with increasing frequency and eventually comes to a halt beyond some critical frequency $\omega_c \simeq 5.5$, which may serve as a first estimation of the critical energy of the localization transition (mobility edge) in our system \cite{Beltukov-2013}. The energy carried by excitations with frequencies below the mobility edge $\omega_c$ spreads in space diffusively with a diffusion coefficient $D(\omega)$, whereas the spread of the energy of excitations with frequencies above the mobility edge slows down with time and eventually stops in the limit of long times, due to Anderson localization. Diffusion coefficient cannot be defined at these frequencies and the localized excitations are characterized by their localization length $\xi(\omega)$.

Displacements of atoms in a harmonic system can be represented as superpositions of eigenmodes $e_i(\omega_n)$:
\begin{equation}
    u_i(t) = \sum_n \frac{v_n}{\omega_n}e_i(\omega_n)\sin(\omega_nt),
\end{equation}
where the initial velocity for each eigenmode is given by
\begin{equation}
    v_n = \sum_i \dot{u}_i(0)e_i(\omega_n).
\end{equation}
However, a direct analysis of eigenmodes $e_i(\omega_n)$ via diagonalization of the dynamical matrix $\hat{M}$ requires unreasonably big computational resources growing rapidly with the matrix size $N$. Therefore, in order to analyze the behavior of excitations as a function of their frequency $\omega$, we perform a windowed Fourier transform
\begin{equation}
    u_i(\omega, t) = 2\int\limits_{-\infty}^{\infty} u_i(t-t') W(t')\cos(\omega t')dt',
\end{equation}
where $W(t)$ is a symmetric window function with a normalization $2\pi \int_{-\infty}^{\infty} W^2(t)dt = 1$. For negative times $t$, $u_i(t)$ is defined as $u_i(-t)=-u_i(t)$, reflecting the time-reversal symmetry of the Newton equation (\ref{eq:newton}). Because Eq.~(\ref{eq:newton}) is linear, $u_i(\omega, t)$ is its solution for any given frequency $\omega$. Therefore, $u_i(\omega, t)$ can be considered as a quasimonochromatic excitation of frequency $\omega$.

The energy $E(\omega)$ of the quasimonochromatic excitation $u_i(\omega, t)$ is an integral of motion for each frequency $\omega$. It can be presented as a sum of partial energies carried by eigenmodes:
\begin{equation}
    E(\omega) = \sum_n \left[ K(\omega-\omega_n) + K(\omega+\omega_n) \right] E_n,
\end{equation}
where $E_n = v_n^2/2$ is a partial energy carried by $n$-th eigenmode in the broadband excitation $u_i(t)$. The function $K(\omega)$ is a broadened delta-function which selects a narrow frequency interval near $\pm \omega$. Since we consider positive frequencies only, the term $K(\omega+\omega_n)$ can be neglected. The selective function $K(\omega)$ depends on the window function $W(t)$ as
\begin{equation}
   K(\omega) = \left[ \int\limits_{-\infty}^{\infty} W(t) \cos(\omega t)\,dt \right]^2.
\end{equation}
We use the window function of the form $W(t) = (2\pi\tau)^{-1/2}\cos(\pi t/2\tau)$ if $|t| < \tau$ and $W(t) = 0$ otherwise. In this case, $K(\omega) = 2\pi \tau \cos^2(\omega\tau)/(\pi^2 + 4\omega^2\tau^2)^2$. We set the half-width of the window function $\tau = 60$, which ensures a good enough frequency resolution for the further numerical analysis.

The energy of the quasimonochromatic excitation localized on the $i$-th atom can be written as
\begin{equation}
    E_i(\omega, t) =  \frac{1}{2}\dot{u}_i(\omega, t)^2 + \frac{1}{2}\sum_{j}M_{ij}u_i(\omega, t)u_j(\omega, t),
\end{equation}
whereas the one-dimensional energy density is
\begin{equation}
    \phi(\omega, x,t) = \sum_i E_i(\omega, t)\delta(x - x_i).
\end{equation}

The square of the penetration depth of the initial excitation into the right half of the sample can be defined as
\begin{equation}
    R^2(\omega, t) = \frac{1}{\phi_0(\omega)}\int\limits_0^{\infty} x \phi(\omega, x,t) dx,   \label{depth}
\end{equation}
where $\phi_0(\omega)$ is the average energy density in the left half of the system at $t=0$. The average energy density can be found as $\phi_0(\omega)=2E(\omega)/L_x$ where $L_x$ is the width of the system in $x$ direction. The squared penetration depth $R^2(\omega, t)$ is averaged over several (typically 10) realizations of disorder to obtain an ensemble-averaged quantity $\langle R^2(\omega, t) \rangle$ which is used as a starting point of the finite-time scaling analysis described in the next section.

\begin{figure}
\includegraphics[width=0.98\columnwidth]{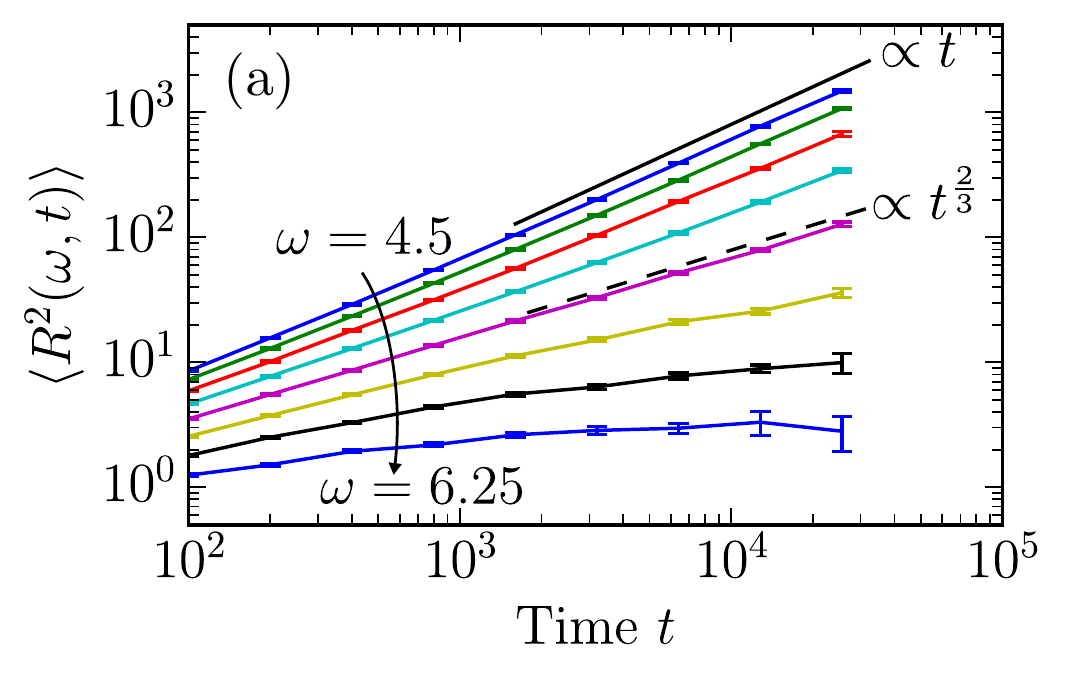}
\includegraphics[width=0.98\columnwidth]{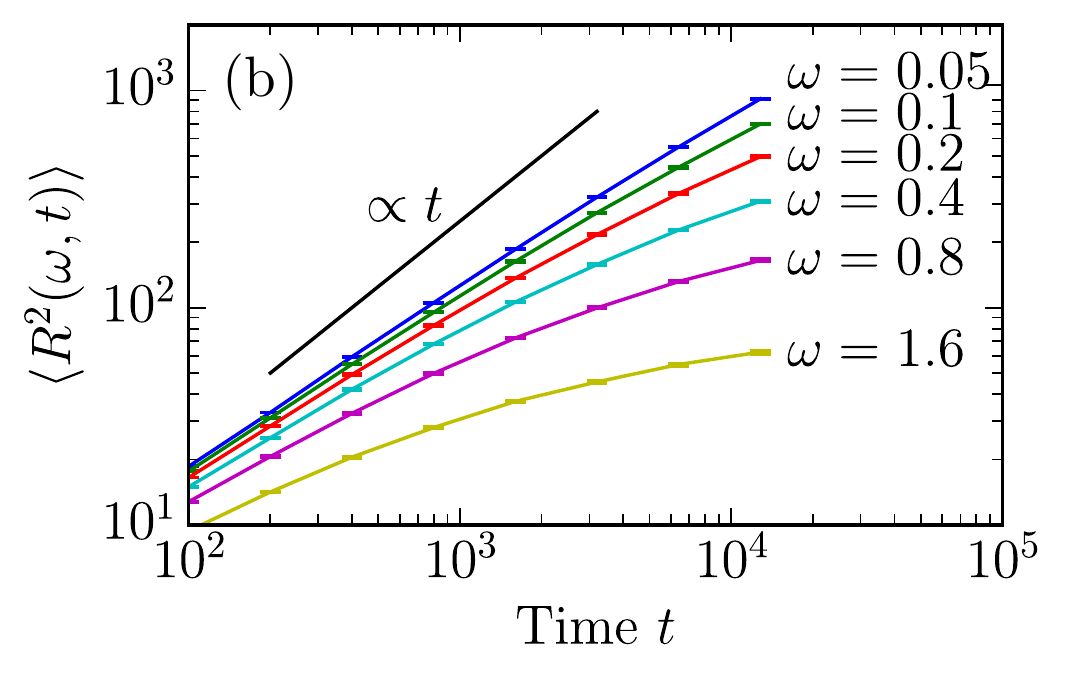}
\vspace*{-2mm}
\caption{\label{fig_r2} (a) Average square of the penetration depth $\langle R^2(\omega,t) \rangle$ for a set of frequencies ($\omega = 4.5$, 4.75, 5, 5.25, 5.5, 5.75, 6 and 6.25 from top to bottom) around the localization transition of a 3D system. The solid and dashed straight lines illustrate $\langle R^2 \rangle \propto t$ for $\omega < \omega_c \simeq 5.5$ and $\langle R^2 \rangle \propto t^{2/3}$ expected at the mobility edge $\omega = \omega_c$, respectively. (b) The same for a 2D system.}
\end{figure}

\section{Finite-time scaling analysis}
\label{sec_fts}

In the following, we will be interested only in the behavior of our disordered system in the long-time limit (ideally, $t \to \infty$). Figure \ref{fig_r2}(a) illustrates the time-dependence of $\langle R^2(\omega, t) \rangle$ in different regimes of wave propagation corresponding to different frequencies $\omega$. In the diffuse regime of propagation ($\omega < \omega_c$), the ensemble-averaged energy density obeys a diffusion equation and we find $\langle \phi(\omega,x,t) \rangle = [\phi_0(\omega)/2] \operatorname{erfc}(x/\sqrt{4D(\omega)t})$, yielding $\langle R^2(\omega,t) \rangle = \frac12 D(\omega) t$. On the other hand, $\langle R^2(\omega,t) \rangle$ saturates at a value of the order of the square of the localization length $\xi(\omega)^2$ when Anderson localization sets in ($\omega > \omega_c$). At the mobility edge ($\omega = \omega_c$) one expects $\langle R^2(\omega,t) \rangle \propto t^{2/3}$ ~\cite{shapiro82}. For comparison, Fig.\ \ref{fig_r2}(b) shows results that we obtained for a two-dimensional (2D) system where no localization transition is expected and the growth of $\langle R^2(\omega,t) \rangle$ with time $t$ is sublinear with a tendency to saturation at long times, for all frequencies $\omega > 0$.

It is convenient to analyze a ratio $\langle R^2(\omega,t) \rangle/t^{2/3}$ which is expected to be independent of time at the mobility edge. We thus introduce a scaling function
\begin{eqnarray}
F(\omega,t) &=& \rho(\omega)^{2/3} \frac{\langle R^2(\omega,t) \rangle}{t^{2/3}},
\label{fscaling}
\end{eqnarray}
where the density of states $\rho(\omega)$ is introduced to make $F$ dimensionless. In the diffuse regime of transport ($\omega < \omega_c$), $F(\omega, t)$ grows with time $t$, whereas at frequencies $\omega > \omega_c$ at which eigenstates are localized, we expect $F(\omega, t)$ to be a decreasing function of $t$. At the critical point $\omega=\omega_c$, $\langle R^2(\omega,t) \rangle \propto t^{2/3}$ ~\cite{shapiro82,berk90} and $F(\omega_c, t) = \mathrm{const}$ independent of time. It is instructive to define an effective length scale $L(\omega, t) = [t/\rho(\omega)]^{1/3}$ and rewrite Eq.\ (\ref{fscaling}) as
\begin{eqnarray}
F(\omega,t) &=& \frac{\langle R^2(\omega,t) \rangle}{L^2(\omega,t)}.
\label{fscaling2}
\end{eqnarray}

We now introduce a single-parameter scaling hypothesis (which will be justified \emph{a posteriori} but remains an assumption at this point) that consists in assuming that in the vicinity of localization transition, $F$ is a function of a single parameter $L/\xi$. Here $\xi = \xi(\omega)$ denotes the localization length for $\omega > \omega_c$ and the correlation length of critical fluctuations for $\omega < \omega_c$. It diverges at the transition as $\xi(\omega) \propto (\omega-\omega_c)^{-\nu}$, where $\nu$ is the critical exponent of the transition. The single-parameter scaling hypothesis can be conveniently reformulated in terms of a new variable $\psi = L^{1/\nu} (\omega - \omega_c)/\omega_c \propto (L/\xi)^{1/\nu}$: $F = F(\psi)$. The procedure of finite-size (or finite-time, in our case) scaling consists in estimating the critical parameters of the localization transition ($\omega_c$ and $\nu$) from the best fits of the above expression to numerical (or experimental) data. The finite-size scaling analysis is extensively used in the theoretical studies of Anderson transition; a recent review of related research can be found in Ref.\ \onlinecite{slevin14}. In contrast, the finite-time scaling has been applied to the localization problem only recently \cite{lemarie09}.
Several improvements of the basic scaling procedure described above have been proposed: the dependence of $\psi$ on $w = (\omega - \omega_c)/\omega_c$ may be nonlinear and a dependence of $F$ on an additional, irrelevant scaling variable that is supposed to account for weak deviations from the single-parameter scaling, may be introduced \cite{slevin14}. The final definitions of functions to be used for fits to numerical data are then the following:
\begin{eqnarray}
\psi_i &=& f_i(w) L^{\alpha_i},
\label{psi}
\\
f_i(w) &=& \sum\limits_{j=1}^{m_i} b_{ij} w^j,
\label{taylor1}
\\
\ln F(\psi_1, \psi_2) &=& \sum\limits_{j_1=0}^{n_1} \sum\limits_{j_2=0}^{n_2} a_{j_1 j_2} \psi_1^{j_1} \psi_2^{j_2},
\label{taylor2}
\end{eqnarray}
where $\psi_1 = \psi$, $i = 1, 2$, and the (yet unknown) functions $f_i(w)$ and $\ln F(\psi_1, \psi_2)$ have been expanded in Taylor series. The relevant scaling exponent is $\alpha_1 = 1/\nu$ whereas the irrelevant one is $\alpha_2 = y < 0$. The latter condition ensures that the role of the irrelevant scaling variable $\psi_2$ decreases when the effective ``size'' of the system $L$ (or, equivalently, time $t$) increases. Therefore, the single-parameter scaling is restored in the limit of $L,t \to \infty$.

\begin{figure}
\includegraphics[width=0.98\columnwidth]{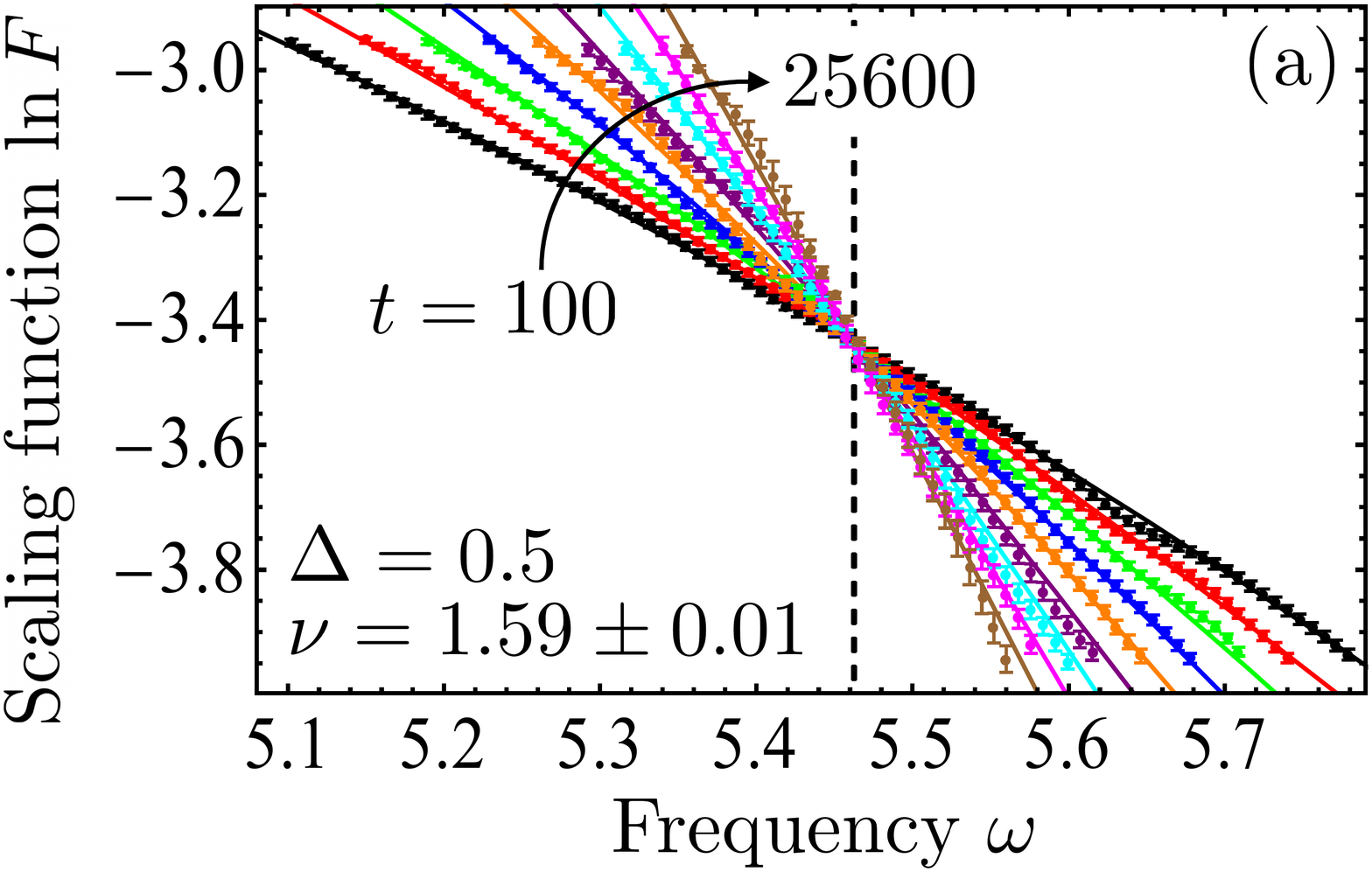}
\includegraphics[width=0.98\columnwidth]{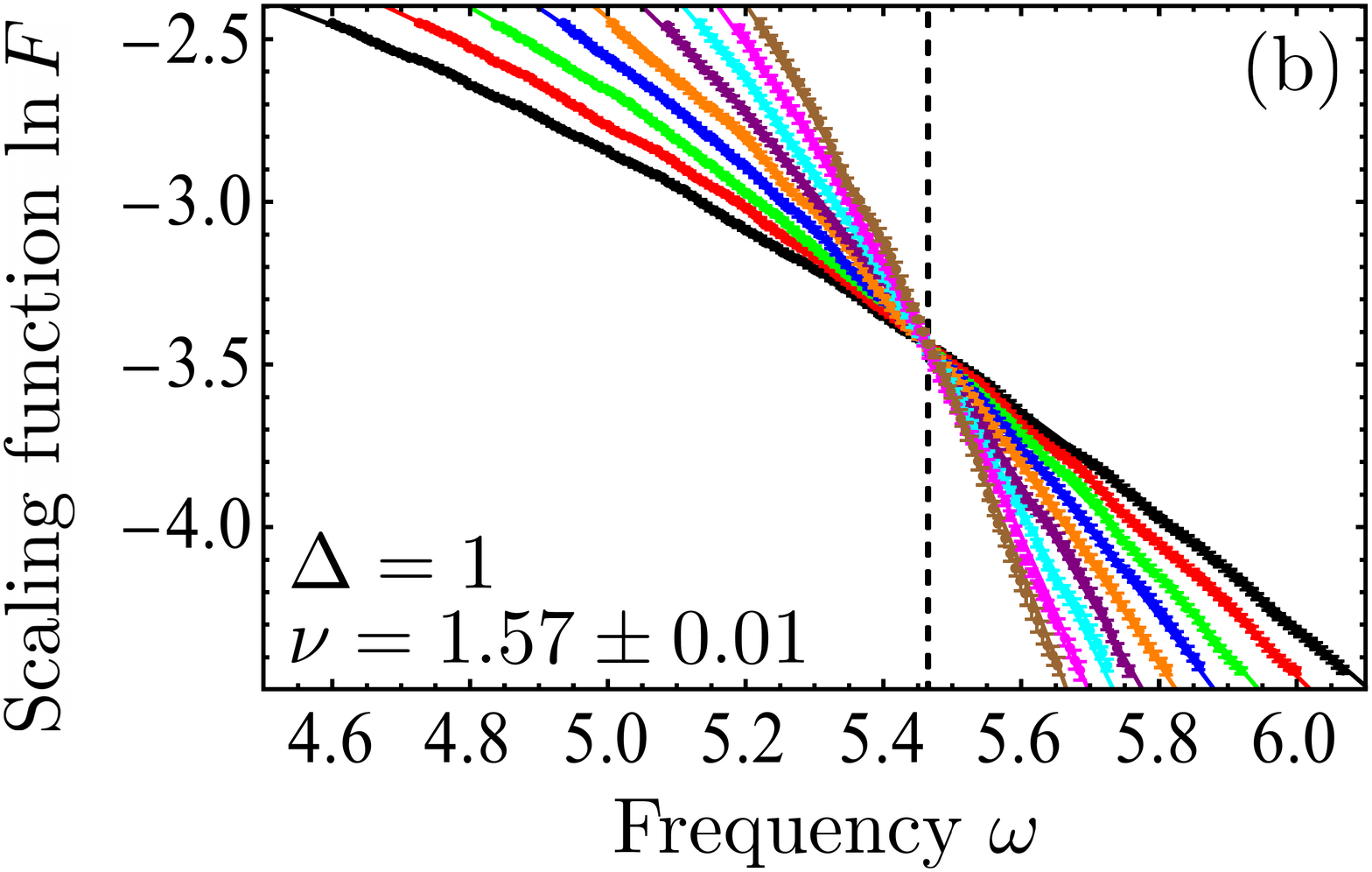}
\vspace*{-2mm}
\caption{\label{fig_best_fits} Best fits to numerical results for two widths of the critical region: $\Delta = 0.5$ (a) and 1 (b). Different colors correspond to different times $t$:
100 (black), 200 (red), 400 (green), 800 (blue), 1600 (orange), 3200 (purple), 6400 (cyan), 12800 (magenta), 25600 (brown). The fits were performed using Eqs.\ (\ref{psi}--\ref{taylor2}) with $m_1 = 2$, $n_1 = 1$, $m_2 = n_2 = 0$ (a) and $m_1 = 3$, $n_1 = 2$, $m_2 = n_2 = 0$ (b). The best-fit values of the mobility edge $\omega_c$ are shown by vertical dashed lines. The best-fit values of the critical exponent $\nu$ are written on the plots.}
\end{figure}

In order to demonstrate the localization transition in our model and estimate its critical parameters, we determine $F(\omega, t)$ using Eq.\ (\ref{fscaling}) from the numerically computed $\langle R^2(\omega,t) \rangle$.  Within numerical errors, dependencies $\ln F(\omega)$ plotted for a set of different times $t$ between 100 and 25600 cross in a single point $(\tilde\omega_c, \ln \tilde F_c)$, which is a clear evidence of critical behavior. We restrict our consideration to the vicinity of the crossing point and fit the data falling in the range $\ln F \in [\ln\tilde F_c - \Delta, \ln\tilde F_c + \Delta]$ to Eqs.\ (\ref{psi}--\ref{taylor2}) (see Appendix \ref{A} for details). To ensure that only the vicinity of the mobility edge is analyzed, we choose $\Delta$ to be small: $\Delta = 0.5$ or 1. The critical parameters of the localization transition---the mobility edge $\omega_c$ and the critical exponent $\nu$---are obtained as the best fit parameters; their uncertainties $\delta\omega_c$ and $\delta\nu$ are also provided by the fitting routine \cite{math} and the fit results are reported as $\omega_c \pm \delta\omega_c$ and $\nu \pm \delta\nu$ in the following. The fits are repeated with different orders $m_i$, $n_i$ of Taylor expansions in Eqs.\ (\ref{psi}) and (\ref{taylor1}), and using only the data corresponding to times $t$ larger than some minimum $t_{\mathrm{min}}$. The quality of the fits is characterized by the $\chi^2$ statistics (see Appendix \ref{A}).

The analysis of fit results shows that all fits yield consistent values of critical parameters. For any $m_1, n_1  \leq 3$, $m_2 \leq m_1$, $n_2 \leq n_1$, $t_{\mathrm{min}} = 100$--1600 and the two values of $\Delta$ used, the best-fit values of $\omega_c$ remain in the range 5.462--5.518 and the best-fit values of $\nu$ in the range 1.214--1.987. However, extreme values in these ranges are obtained either when $m_i$, $n_i$ are not large enough and the fit quality is poor ($\chi^2$ statistics much larger than 1) or when they are too large and Eqs.\ (\ref{psi}--\ref{taylor2}) overfit the data ($\chi^2$ below 1). ``Optimal'' fits that require a minimum number of free fit parameters and, at the same time, yield $\chi^2 \sim 1$, are obtained with $m_1 = 2$, $n_1 = 1$ for $\Delta = 0.5$ and  $m_1 = 3$, $n_1 = 2$ for $\Delta = 1$, without using the irrelevant scaling variable ($m_2 = n_2 = 0$). Introducing the irrelevant variable improves the quality of the fits only slightly and leads to a significant spread of best-fit values of $\omega_c$ and $\nu$ obtained for different $t_{\mathrm{min}}$. We thus judged using the irrelevant scaling variable inappropriate for our numerical data. The optimal fits to all available date ($t_{\mathrm{min}} = 100$) are shown in Fig.\ \ref{fig_best_fits}.

\begin{figure}
\includegraphics[width=0.98\columnwidth]{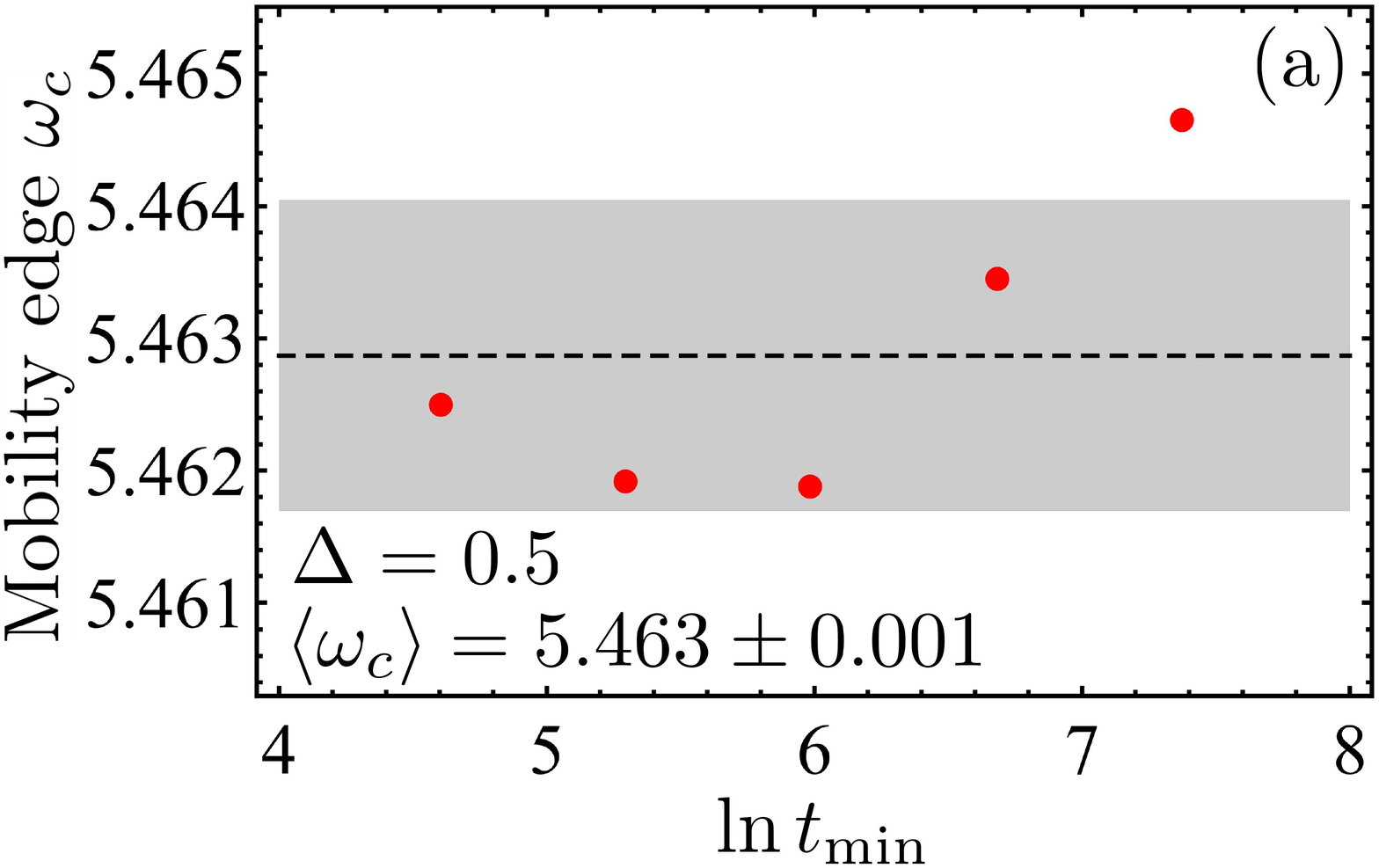}
\includegraphics[width=0.98\columnwidth]{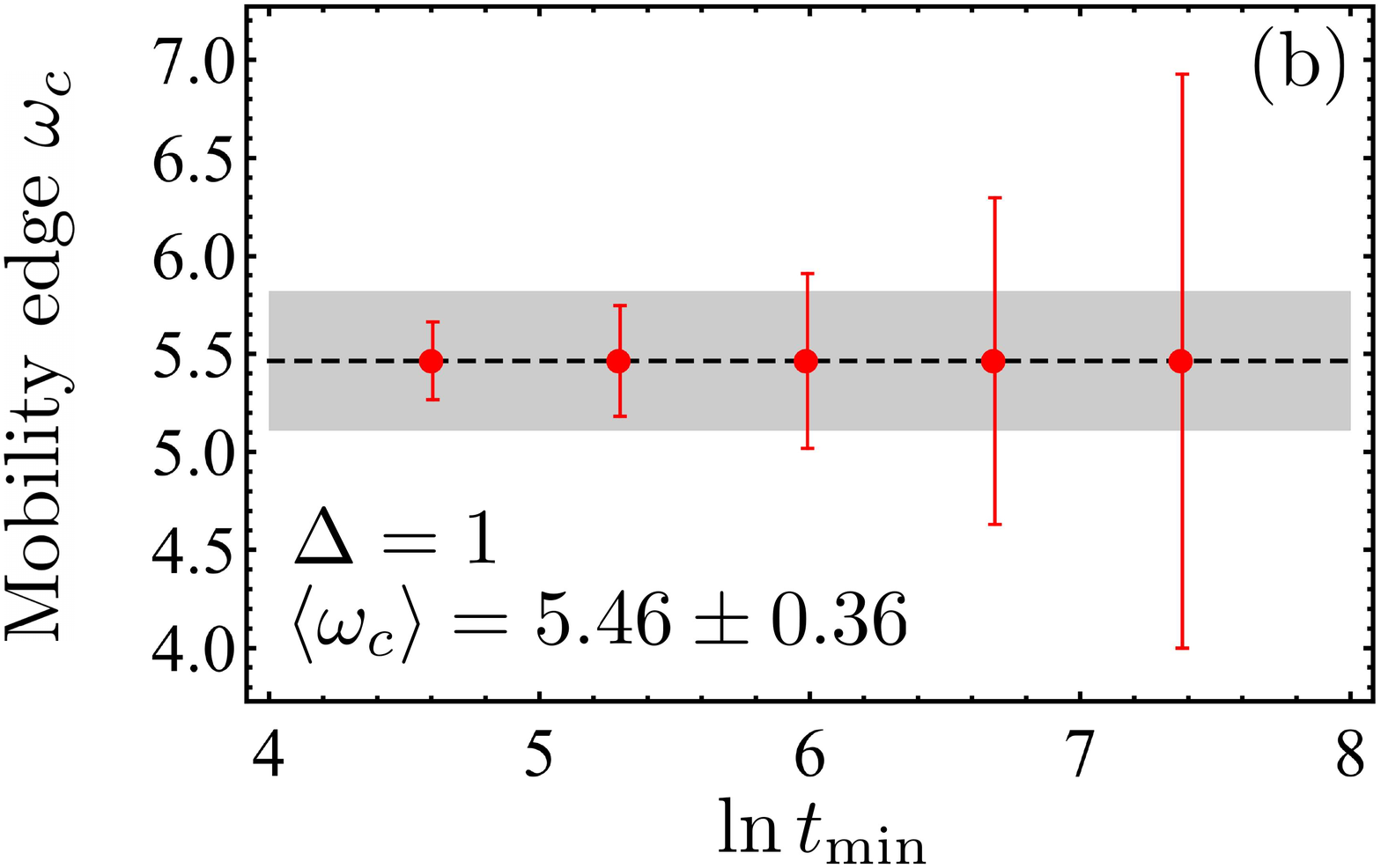}
\vspace*{-2mm}
\caption{\label{fig_wc} Best-fit values of the mobility edge $\omega_c$ as functions of the minimum time $t_{\mathrm{min}}$ used in the fit procedure and for two different widths of the critical region: $\Delta = 0.5$ (a) and 1 (b). Error bars show the standard errors of the best-fit values, the horizontal dashed lines show the values $\langle \omega_c \rangle$ of $\omega_c$ averaged over all $t_{\mathrm{min}}$, the gray shaded regions show the errors of $\langle \omega_c \rangle$.}
\end{figure}

\begin{figure}
\includegraphics[width=0.98\columnwidth]{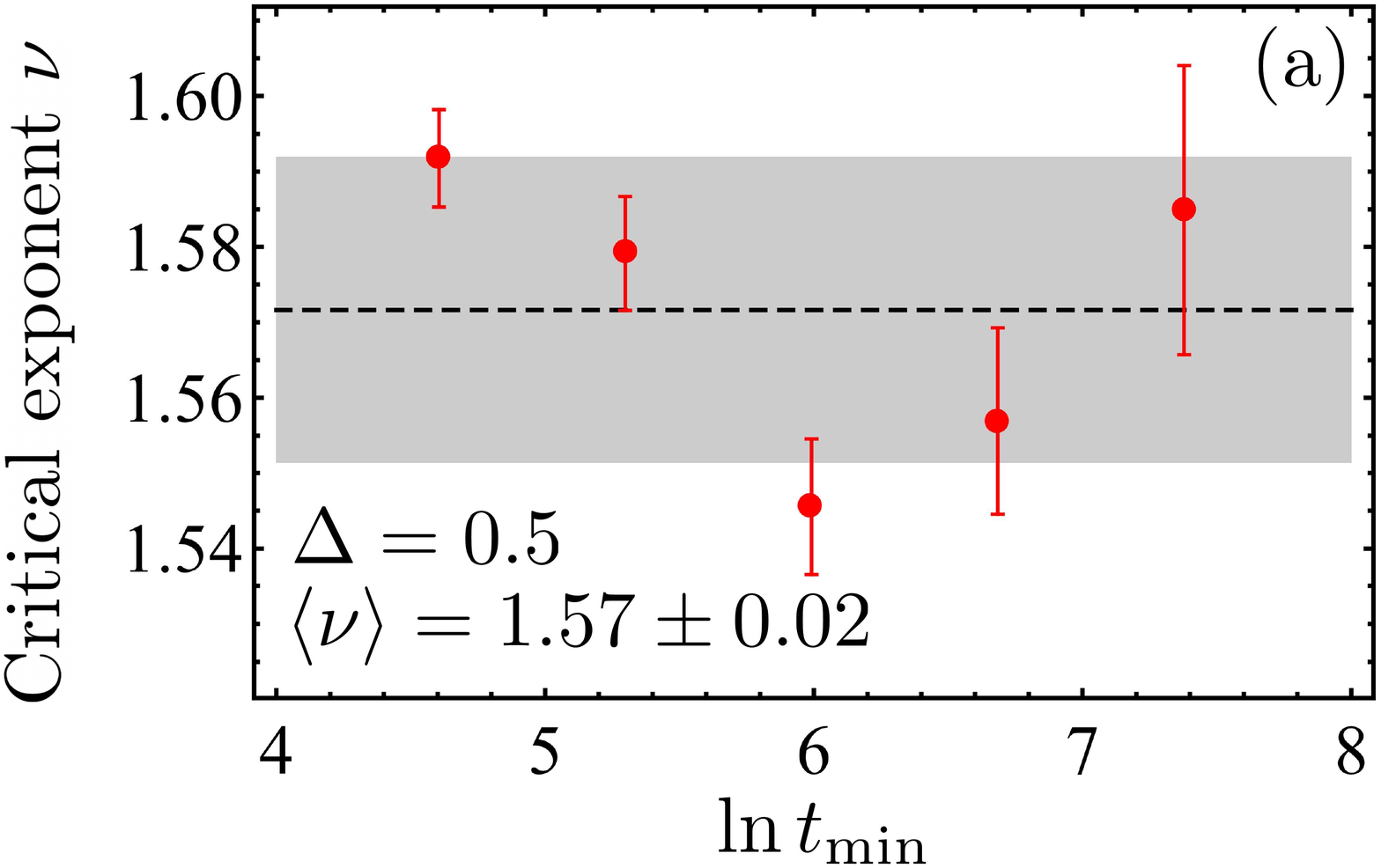}
\includegraphics[width=0.98\columnwidth]{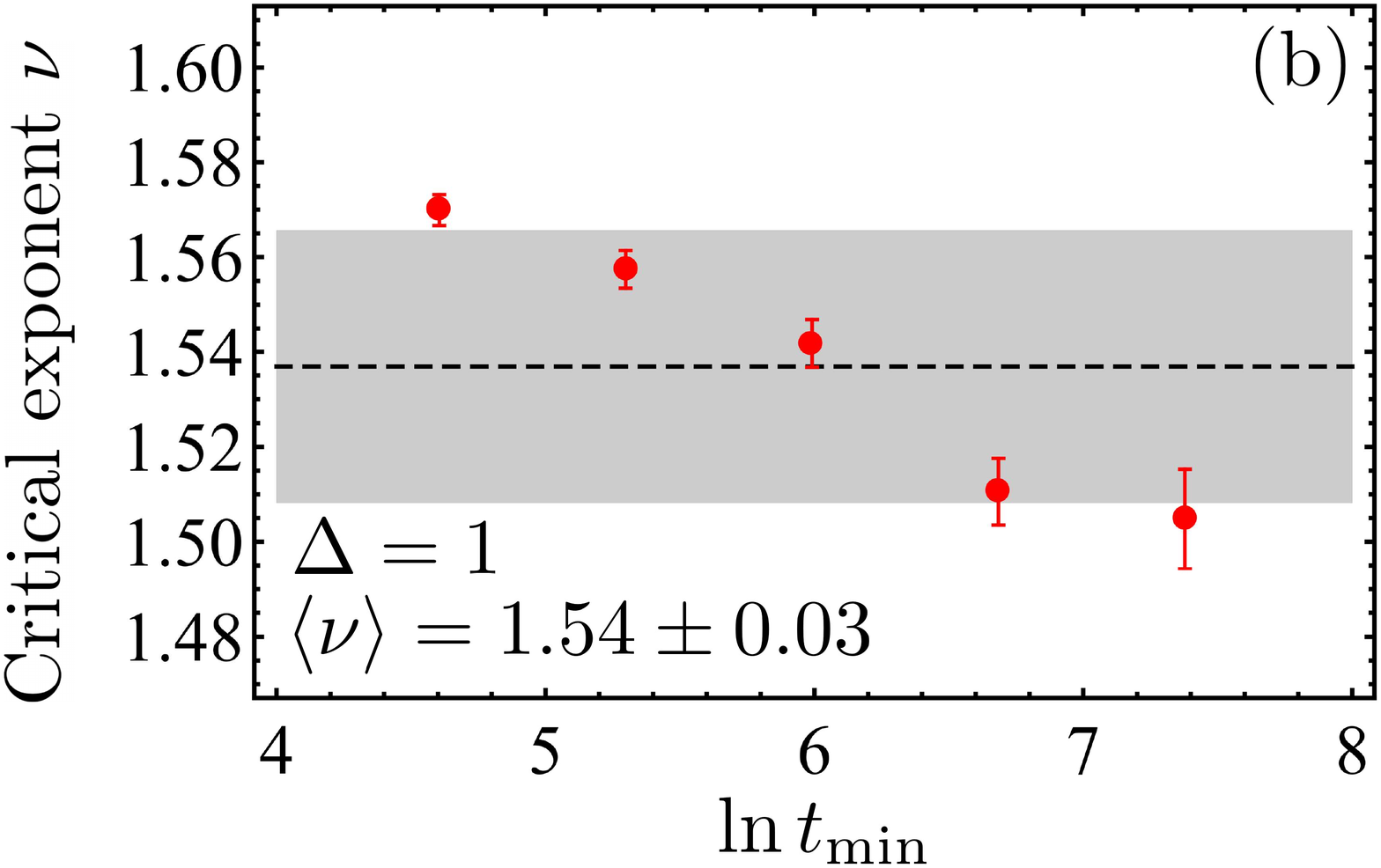}
\vspace*{-2mm}
\caption{\label{fig_v} Best-fit values of the critical exponent $\nu$ as functions of the minimum time $t_{\mathrm{min}}$ used in the fit procedure and for two different widths of the critical region: $\Delta = 0.5$ (a) and 1 (b). Error bars show the standard errors of the best-fit values, the horizontal dashed lines show the values $\langle \nu \rangle$ of $\nu$ averaged over all $t_{\mathrm{min}}$, the gray shaded regions show the errors of $\langle \nu \rangle$.}
\end{figure}

A better idea of the accuracy of the estimates of critical parameters following from the fits can be obtained by analyzing the best-fit values of $\omega_c$ and $\nu$ obtained for different $t_{\mathrm{min}}$. The latter are shown in Figs.\ \ref{fig_wc} and \ref{fig_v}, respectively. We obtain the final estimates of critical parameters $\langle \omega_c \rangle$ and $\langle \nu \rangle$ consistent with all $t_{\mathrm{min}}$ as averages over $\omega_c$ and $\nu$ obtained for different $t_{\mathrm{min}}$. These estimates are given in Figs.\ \ref{fig_wc} and \ref{fig_v}.

\section{Discussion}
\label{sec_disc}

\begin{figure}
\includegraphics[width=0.98\columnwidth]{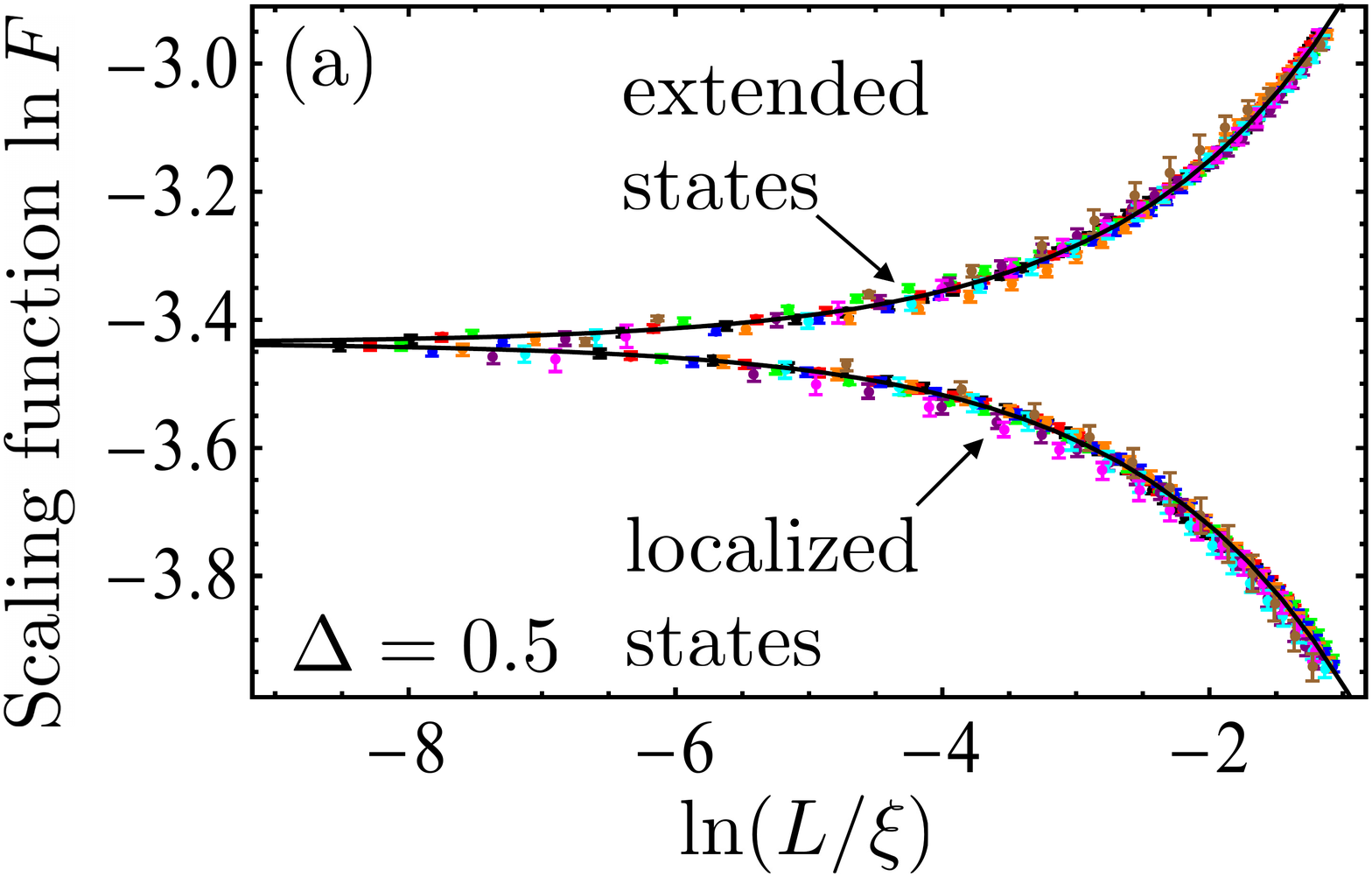}
\includegraphics[width=0.98\columnwidth]{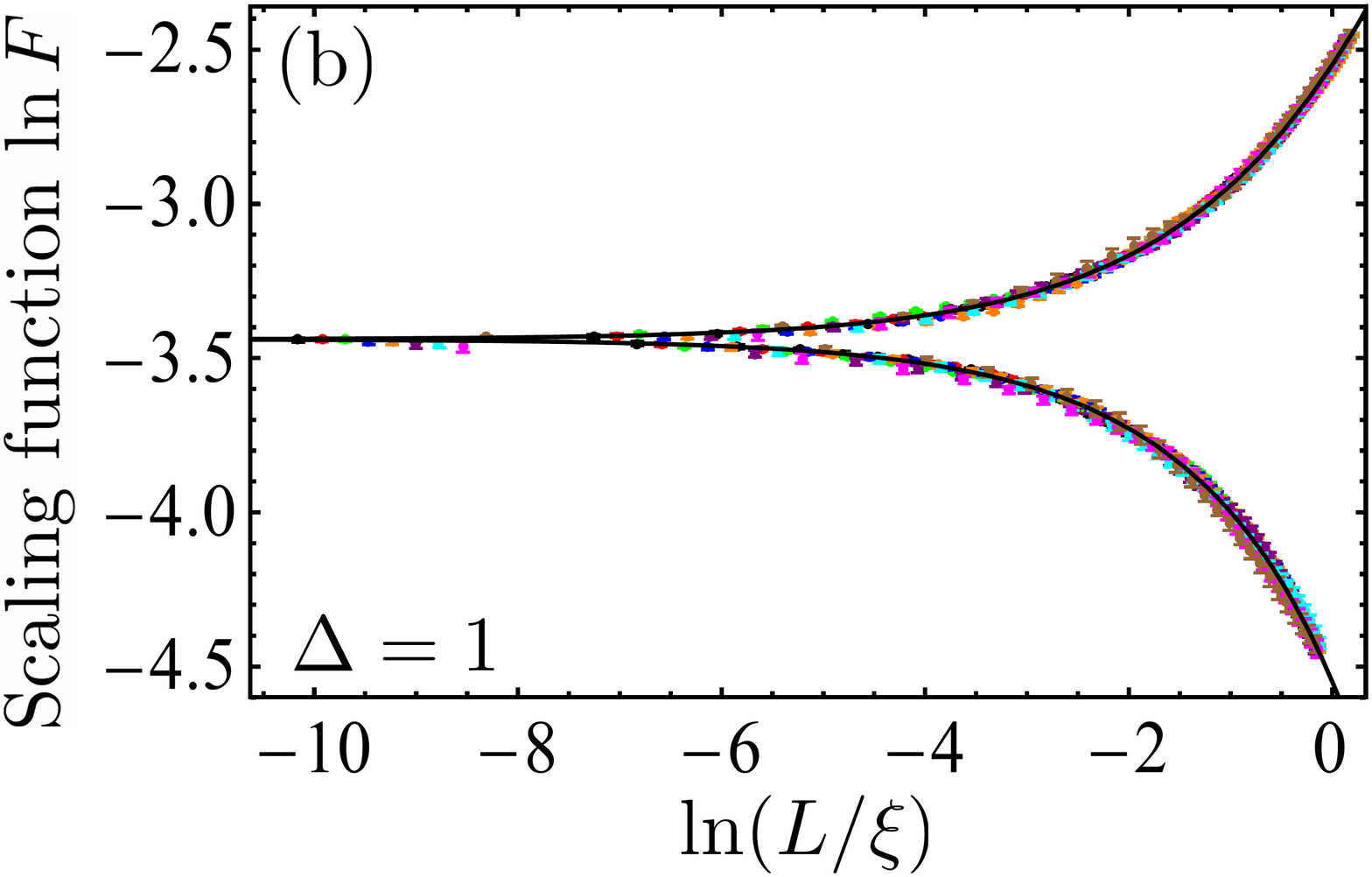}
\vspace*{-2mm}
\caption{\label{fig_sps} Data of Fig.\ \ref{fig_best_fits} represented as a function of a single parameter $L/\xi$. All data points (symbols) fall on a single master curve following from Eq.\ (\ref{taylor2}) (black solid line) justifying the hypothesis of single-parameter scaling.}
\end{figure}

The analysis presented in Sec.\ \ref{sec_fts} was based on an important hypothesis of single-parameter scaling which was assumed without justification. We now can check if our numerical data are consistent with this hypothesis by plotting the data of Fig.\ \ref{fig_best_fits} as a function of a single parameter $L(\omega, t)/\xi(\omega)$, where $\xi(\omega) = \mathrm{const}_{\pm} \times |u_1(w)|^{-\nu}$ (see Fig.\ \ref{fig_sps}). The numerical constants $\mathrm{const}_{\pm}$ cannot be determined from a scaling analysis. Figure\ \ref{fig_sps} clearly confirms that the hypothesis of single-parameter scaling is justified because all points resulting from our numerical analysis fall on a single master curve. The latter has two branches: the first one corresponding to extended states ($\ln F$ growing with $L/\xi$) and the second one corresponding to localized states ($\ln F$ decreasing with $L/\xi$). The cusp at which the two branches join corresponds to the critical point of Anderson transition where $\xi$ tends to infinity and $\ln F$ is independent of $L$.

Let us now verify whether the values of critical parameters: the mobility edge $\omega_c$ and the critical exponent $\nu$ obtained in Sec.\ \ref{sec_fts}, are in agreement with the available well-established results for the 3D Anderson transition \cite{evers08}. First, the mobility edge $\omega_c$ is often believed to follow from the celebrated Ioffe-Regel criterion of Anderson localization: $k(\omega_{\rm IR}) \ell(\omega_{\rm IR}) = 1$, where $k(\omega)$ is the wave number and $\ell(\omega)$ is the scattering mean free path for an excitation of frequency $\omega$ \cite{shengbook}. The Ioffe-Regel frequency $\omega_{\rm IR}$ determined from this criterion is often expected to be a more or less reliable estimation of $\omega_c$. However, the model considered in the present work is quite singular because, as was shown previously in Ref.\ \onlinecite{Beltukov-2013}, it does not have macroscopic rigidity. As a result, some characteristic frequencies---such as, e.g., the Ioffe-Regel frequency $\omega_{\rm IR}$---become equal to zero. The Ioffe-Regel criterion is, therefore, not a good criterion of Anderson localization in our system and the mobility edge $\omega_c$ cannot be estimated from it. It is important to realize that the impossibility of estimating $\omega_c$ from the Ioffe-Regel criterion is not an artifact of having zero rigidity. The latter can be readily introduced in our model in several ways~\cite{Beltukov-2013, Beltukov-2016b} and produces nonzero values of $\omega_{\rm IR}$. However, even in this case $\omega_{\rm IR}$ turns out to be considerably lower than the critical frequency of Anderson transition $\omega_c$. As a matter of fact, $\omega_{\rm IR} \ll \omega_c$ is a usual situation for vibrations in amorphous dielectrics \cite{sheng94,Allen-1999,Taraskin-2000}. Under such conditions, we do not expect the macroscopic rigidity to significantly modify vibrational properties at frequencies $\omega \sim \omega_c$, which justifies the use of a model with a vanishing rigidity in our analysis.

The estimation of the critical exponent $\nu = 1.57 \pm 0.02$ that follows from our calculations \cite{exponent} is in agreement with the values obtained previously for both spinless electrons \cite{slevin14} and mechanical vibrations \cite{pinski12a} in solids using the finite-size scaling approach, as well as with the results of finite-time scaling analysis of the kicked rotator model \cite{lemarie09}. This confirms that the localization transition in the considered vibrational system belongs to the same orthogonal universality class as the transitions in the three systems listed above. Our results also validate the use of the finite-time scaling approach as a valuable alternative to the final-size one in mechanical systems with disorder. Such a validation is particularly important in view of the possible application of finite-time scaling analysis to experimental data.

\section{Conclusions}
\label{sec_concl}

In this work, we have presented a finite-time scaling approach to the study of Anderson localization transition for vibration in 3D solids. Application of the approach to a model system in which vibrations are allowed only along a given direction in space (scalar model) has allowed us to estimate the critical frequency and exponent of the transition. The value of the critical exponent is in agreement with the values that were found previously for different 3D physical systems where localization transitions belong to the same universality class. This validates the use of the finite-time scaling approach instead of the better known finite-size scaling method. Finite-time scaling analysis may be particularly interesting for the analysis of experimental data because obtaining time-dependent data for a single disordered sample may be easier than repeating measurements on an ensemble of statistically equivalent samples of different sizes as would be required for the use of the finite-size scaling analysis. Being now tested on the simplest scalar model of vibrations, our approach can be applied to the analysis of more realistic, vectorial models in which displacements of masses constituting a disordered sample are three-dimensional vectors \cite{skip18}.

\begin{acknowledgments}
The work of S.E.S. is supported by the Agence Nationale de la Recherche (grant ANR-14-CE26-0032 LOVE). The work of Y.M.B. is supported by the Council of the President of the Russian Federation for Support of Young Scientists and Scientific Schools (project no. SP-3299.2016.1). A part of the computations presented in this paper was performed using the Froggy platform of the CIMENT infrastructure ({\tt https://ciment.ujf-grenoble.fr}), which is supported by the Rhone-Alpes region (GRANT CPER07\_13 CIRA) and the Equip@Meso project (reference ANR-10-EQPX-29-01) of the programme Investissements d'Avenir supervised by the Agence Nationale de la Recherche.
\end{acknowledgments}

\appendix

\section{Details of the procedure applied to fit the numerical data}
\label{A}

We provide below the details of the procedure applied to fit the numerical data for the scaling function $F(\omega, t)$. First, we plot $\ln F$ as a function of $\omega$ for a set of different times $t$. We observe that within numerical errors, all lines corresponding to different times cross in a single point $(\tilde\omega_c, \ln \tilde F_c)$. We now restrict our consideration to the vicinity of the crossing point and fit the data falling in the range $\ln F \in [\ln\tilde F_c - \Delta, \ln\tilde F_c + \Delta]$, with $\Delta \lesssim 1$, to Eqs.\ (\ref{psi}--\ref{taylor2}) with fixed $m_i$, $n_i$. To reduce the sensitivity of results to the starting values of fit parameters, we repeat the fitting procedure 100 times for each $\Delta$ and each set of $m_i$, $n_i$ with random starting values of $\omega_c$ within $\pm 20\%$ of $\tilde\omega_c$, $\nu \in [0, 10]$, $y \in [-10, 0]$, $a_{00}$ within $\pm 20\%$ of $\ln\tilde F_c$, and $b_{ij}$, $a_{j_1 j_2}$ in a wide range $[-10, 10]$ (except for $b_{10} = 0$ and $a_{01} = a_{10} = 1$, see Ref.\ \onlinecite{slevin14}). The fits are accepted if the best-fit value of $\omega_c$ falls within $\pm 20\%$ of $\tilde\omega_c$, $y < -1$ and the contribution of the irrelevant scaling variable does not exceed 10\%. Among the accepted fits, the best fit is chosen as the one having the minimum value of the $\chi^2$ statistic:
\begin{eqnarray}
\chi^2 = \frac{1}{N_{\mathrm{data}}} \sum\limits_{i = 1}^{N_{\mathrm{data}}}
\frac{\left[ \ln F_i^{\mathrm{(data)}}-\ln F_i^{\mathrm{(fit)}} \right]^2}{\sigma_i^2},
\label{chi2}
\end{eqnarray}
where $N_{\mathrm{data}}$ is the total number of data points corresponding to different times and frequencies, the superscripts `(data)' and `(fit)' denote the values of $\ln F$ obtained from the data using Eq.\ (\ref{fscaling}) and from the fit function (\ref{taylor2}), respectively. $\sigma_i$ is the statistical error of $\ln F_i^{\mathrm{(data)}}$.



\begin{thebibliography}{99}
\bibitem{loc2d}
For disordered Schr\"{o}dinger and scalar Helmholtz equations, all modes are expected to be localized in 1D and 2D for an arbitrary weak disorder with short-range spatial correlations, provided that disorder preserves the time-reversal symmetry and the spin-rotation invariance \cite{abrahams79}.
\bibitem{exponent}
The critical exponent determines the behavior of the system in the immediate vicinity of the localization transition and its estimation from the analysis of a narrower region around the critical point is likely to be more accurate. That is why we consider the estimation obtained for $\Delta = 0.5$ as the best result.
\end{thebibliography}
\end{document}